# Angular dependent resonant photoemission processes at the 2*p* thresholds in nickel metal

M. Magnuson, A. Nilsson, M. Weinelt and N. Mårtensson

*Department of Physics, Uppsala University, P. O. Box 530, S-751 21 Uppsala, Sweden*

**Abstract**

Angle-resolved valence-band resonant photoemission of nickel metal has been measured close to the 2*p* core-level thresholds with synchrotron radiation. The well-known 6-eV correlation satellite has an intensity enhancement of about two orders of magnitude at resonance. The angular dependence of the photoemission intensity has been studied as function of photon energy and provides unambiguous evidence for interference effects all the way up to the resonance maximum. The observation of different angular asymmetries, β, for the valence band and the satellite is discussed in connection to the origin of the resonant photoemission process and the character of the satellite.

## 1 Introduction

The phenomenon of resonant photoemission in correlated systems has been extensively discussed over the years, see e. g. [1,2]. A massive intensity enhancement of the emission cross section is observed when, in addition to the direct photoemission channel, a second core-assisted Auger-like channel is being opened. The important question in the context of a true resonant photoemission process is whether the direct and indirect channels are added coherently or incoherently. For resonant photoemission, the amplitudes of the individual processes are added. This leads to interference terms in the expression for the intensity which depend on both processes. On the contrary, if the process is incoherent the resulting intensity is obtained as the sum of the intensities of the two individual contributions without any interference terms. The interference effects lead to the largest redistribution of intensity when the two participating channels are of similar strength. Thus, a massive intensity increase does not necessarily imply clearly visible interference effects.

In connection to resonant photoemission, Ni metal has been extensively discussed [3,4,5,6,7,8]. One way to find out if there are any signatures of interference terms between the direct photoemission and the core-hole assisted process is to measure the energy dependence of the photoemission cross section. A strong intensity enhancement of the well-known 6-eV satellite in Ni metal leads to a Fano profile [9] of the photoemission cross section which has been observed both at the 3*p* and 2*p* core-level thresholds [10, 11]. The shape of a Fano profile is determined by the asymmetry parameter, q, under the assumption that there is an interaction between one discrete state and a continuum. In Ref [11], a very strong intensity enhancement (q ≥ 9) was observed for the 6-eV satellite. For the valence band a much smaller q-value was found (q ~ 1.5). The spectral profiles could not be well fitted within the simple Fano model since the interaction probably involves several discrete states and continua [11]. Furthermore, it was observed that the profiles were different in two angular orientations giving strong evidence for interference effects.

To gain further insight into the properties of the resonant process, we have performed angle-resolved measurements to obtain the angular asymmetry for excitation energies around the Ni 2*p* core-level threshold. By changing the direction of the polarization vector of the incoming photon beam, the relative strengths of the matrix elements of the individual processes can be varied. If the resonant photoemission process would simply be an incoherent superposition of the two channels, the angular asymmetry parameter β would be a straighforward weighted average of those of the direct photoemission and Auger channels with the weight factors being directly related to the relative intensities of the individual processes. Recently, these kind of measurements have been applied to Ni metal and other systems by Lopez *et al.* [12,13,14,15,16,17], where the direct photoemission process was assumed to have a $β_{PE}$ -value of 2, corresponding to a $cos^2$ angular dependence while the Auger signal was expected to be completely isotropic corresponding to $β_A$ = 0. However, the resolution and energy tuning range did not allow to obtain detailed values of the β-parameter around the threshold resonances [18]. We have followed these ideas and performed detailed spectroscopic studies at high resolution and with accurate energy tuning.

In the present paper we investigate the angular dependence of the valence band photoemission process in Ni at photon energies around the 2*p* edges. The 2*p* edges give better opportunities for studies of resonance effects than the more shallow and broader 3*p* edges since the resonances are more separated by the larger spin-orbit splitting. The resonance behaviour of the 6-eV satellite at the 2*p* edges is found to be very strong with intensities about two orders of magnitude larger than in the non-resonant case. As will be shown, the angular asymmetry parameters $β_{PE}$, of the valence band and the satellite in the photoemission spectrum strongly depend on the excitation energy. The details of the photon energy dependence show that the strong photoemission intensity enhancement at threshold is due to a coherent superposition of the direct photoemission and Auger-like channels.

## 2 Experimental Details

The experiments were performed using monochromatized synchrotron radiation at beamline 8.0 at *the Advanced Light Source*, Lawrence Berkeley National Laboratory. The beamline includes a 5 cm period undulator and a spherical-grating monochromator. The base pressure was better than $2×10^{−10}$ Torr during preparations and measurements. The sample was a Ni(100) single crystal which was cleaned by means of cyclic argon-ion bombardment and annealing in order to remove surface contaminants. For normalization purposes, the photon flux was continuously monitored using a gold mesh in front of the sample.

The incident radiation impinged on the sample at about 5 degrees grazing incidence with respect to the sample surface with the polarization vector at different angles relative to the surface plane. The photoemission spectra were measured with a Scienta SES200 spectrometer [19] with an energy resolution at the Ni 2*p* edge of approximately 0.15 eV and with a monochromator resolution of 0.15 eV. In order to avoid diffraction effects in the spectral intensity, the spectra were always recorded at normal emission with respect to the sample surface, and the polarization vector of the incoming photons was changed by simultaneously rotating both the sample and the spectrometer.

## 3 Results and Discussion

Figure 1 shows on a kinetic energy scale, a set of valence-band photoemission spectra of Ni measued at different excitation energies up to and above the $L_3$ absorption threshold at 852.7 eV. The excitation energies are indicated by the arrows in the absorption spectrum in the inset and correspond to the Fermi level cut-off in each spectrum. The intensities were normalized to the incoming photon flux. The dashed and solid curves correspond to measurements with the **E**-vector in the plane of the surface and parallel to the surface normal, respectively. The angular rotation of the **E**-vector changes the relative weights of the direct photoemission and Auger matrix elements. The spectra are shown after subtraction of an integral Shirley background [20,21].
Starting with the spectra measured well below resonance we observe two types of final states; valence-band states within ~ 2.3 eV from the Fermi level which dominate the spectra and split-off atomic-like $3d^8$ satellite states at 6 eV from the Fermi level. Far below and above the resonance, the intensities of both spectral features are much larger when the **E**-vector is parallel to the surface normal of the sample, and we denote this the 'photoemission geometry' (solid curves). As the excitation energy approaches the $L_3$ resonance maximum, the relative weight of the satellite states increases. For excitation energies closely below the $L_3$ threshold, the intensity of the 6-eV satellite is lowest in the photoemission geometry [22]. At resonance, both the $3d$-band and the satellite contributions are enhanced and the intensity is almost the same in both geometries, it is only slightly higher in the photoemission geometry. The intensity enhancement at the resonance in the photoemission geometry is a factor of ~ 2.3 and ~ 60 for the valence band and the satellite, respectively. Thus, the resonance behaviour is predominantly affecting the satellite final states. Above resonance, the $L_3M_{45}M_{45}$ Auger peak rapidly appears and stays at the same kinetic energy for excitation energies above resonance. The nonresonant Auger intensity is essentially the same for the two experimental geometries. We denote the geometry with the lowest direct photoemission intensity as the 'Auger geometry' (dotted curves). The intensity enhancement at the resonance in the Auger geometry is a factor of ~ 9.5 and ~ 180 for the valence band and the satellite, respectively. The $L_3M_{45}M_{45}$ Auger peak maximum on the $L_3$ resonance excited at 852.7 eV is 4 times higher than for the regular non-resonant $L_3M_{45}M_{45}$ Auger peak excited at 864.0 eV. When the excitation is non-resonant, far above or below the $2p$ thresholds, the direct photoemission channel from the valence band produces mainly $[3d^94s]^{-1}$ final states while the 6-eV satellite corresponds to $3d^8[4s^2]$ final states. The square brackets indicate that the valence electrons are in a metallic state. On resonance, the Auger (autoionization) channel produces the same type of $3d^8[4s^2]$ final states as for the 6-eV satellite with high intensity while the contribution to the $3d^9[4s]^{-1}$ valence-band final states is much smaller. In resonant photoemission there is an interplay between these Auger-like (autoionization) processes and the direct photoemission [5,11]. One of the problems is to normalize the spectra at different photon energies and different angles to each other. The following procedure was adopted: Auger spectra where the $L_3$ and $L_2$ peaks appeared simultaneously were measured well above the thresholds at different angles. After background subtraction, the same angular dependence for the $L_2$−$M_{45}M_{45}$ and the $L_3$−$M_{45}M_{45}$ Auger peaks was observed. In the two-step description, the angular distribution of the Auger electrons are described by the product $\beta_A = \mathcal{A} \times C_A$. The alignment parameter, $\mathcal{A}$, is due to the $2p \rightarrow 3d$ dipole photoionization and describes the alignment of the ion after the excitation step. The Auger decay parameter, $C_A$, describes the asymmetry of the second step i.e., the Auger transition, which are those of the Coulomb matrix elements including the relative phases. For Auger decay from a specific core-excited state, the $C_A$-parameter is assumed to be photon energy dependent in the two-step approximation. However, the fact that the alignment ($\mathcal{A}$-value) for the $L_2$ (J=1/2) photoionized state is zero implies that the angular distribution is indeed isotropic in this case [2]. This has also been experimentally confirmed for the $L_2$−$M_1M_1$ Auger decay in atomic magnesium [30]. The identical angular dependence of the nonresonant $L_3$−$M_{45}M_{45}$ Auger spectra then implies that also this emission is isotropic for the non-resonant spectra. This provides a means to normalize the spectra above resonance to each other. For each angle the spectra at different photon energies below and above the $L_3$ threshold were then normalized to each other based on the measured photon flux.

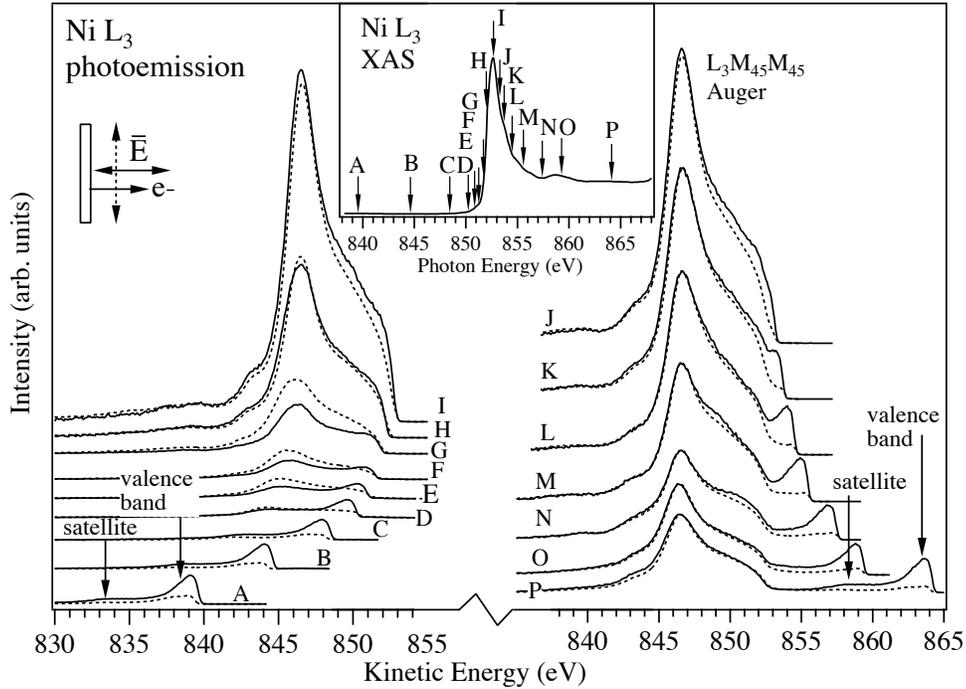

**Fig. 1:** A set of resonant photoemission spectra of Ni metal recorded at different excitation energies close to the $L_3$ threshold with the polarization vector at two different geometries (see text for details). The spectra were excited at the energies indicated by the arrows in the x-ray absorption spectrum in the inset. The kinetic energies are referred to the Fermi level.

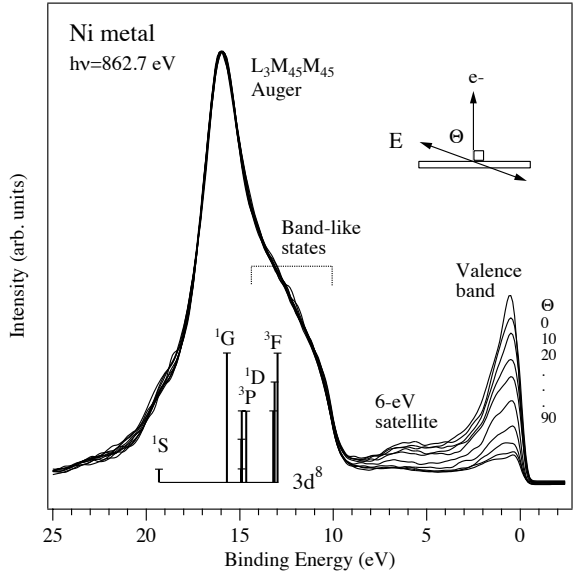

**Fig. 2:** A series of nonresonant valence-band photoemission spectra of Ni recorded at 862.7 eV excitation energy for different angles of the polarization vector. The peak intensities of the $L_3M_{45}M_{45}$ Auger contribution were normalized to the same value.

Figure 2 shows a series of nonresonant valence-band photoemission spectra of Ni metal on a binding energy scale recorded at 862.7 eV as a function of the angle $\tau$, where $\tau$ is the angle between the surface normal and the polarization vector of the incident light. The angle $\tau$ was varied in 10 degree steps between the 'Auger'- and 'photoemission'- geometries corresponding to 90 and 0 degrees, respectively. The spectra in Fig. 2 consist of two main features; the direct valence-band photoemission part observed at 0-10 eV binding energy and the dominating $L_3M_{45}M_{45}$ Auger peak at 15.5 eV 'binding energy'. Both for the Auger and the photoemission channels we observe two types of final states [23,24]. In the Auger spectra we observe both band-like final states (see Fig. 2) and split-off atomic-like $3d^8$ final states. The atomic-like $3d^8$ multiplets consists of the $^1G$ term at the main peak and the $^3F$, $^3P$ and $^1D$ terms which overlap with the band-like states, while the $^1S$ term can be identified on the high binding-energy side of the main Auger peak. In the direct photoemission channel (0-10 eV), we observe similar types of final states as in the Auger peak but in this case the valence-band emission is the dominant channel and the $3d^8$ final states represented by the 6-eV correlation satellite gives a smaller contribution. The angular distribution of the differential photoionization cross section for linearly polarized photons in the dipole approximation can be expressed as [25,26,27];

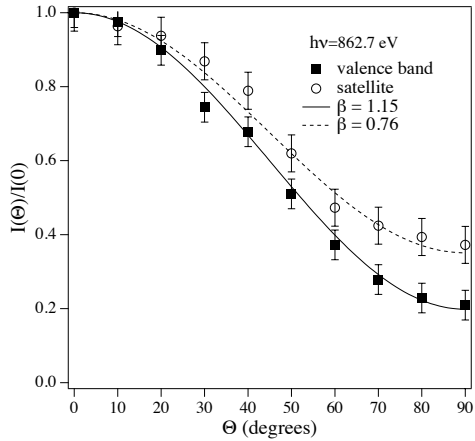

**Fig. 3:** Angular asymmetries of the intensities of the valence band and the satellite using the data from Fig. 2.

$$\frac{d\sigma}{d\Omega} = \frac{\sigma_o}{4\pi}\left[1 + \frac{\beta}{2}(3\cos^2\tau - 1)\right] \qquad (1)$$

where $\sigma_o$ is the absolute photoelectron cross section for a certain excitation energy, $\beta$ is the angular asymmetry parameter and $\tau$ is the angle between the **E**-vector and the outgoing electron. $\beta = +2$ corresponds to a pure $\cos^2\tau$ behaviour, whereas $\beta = 0$ corresponds to an isotropic distribution.

Figure 3 shows the angular asymmetries of the intensities of the valence band photoemission signal and the satellite using the spectra in Fig. 2. The intensities are normalized to the intensity maxima when the angle $\tau$ is 0 degrees, in the photoemission geometry. The full and dotted lines in Fig. 3 represent least-square fits to equation (1). Starting with the valence band, it is interesting to note that the angular intensity distribution does not exhibit a pure $\cos^2$ behaviour corresponding to a $\beta_{PE}$-value of 2. Instead, the $\beta_{PE}$-value is found to be about 1.15 (solid curve). The anisotropy of the photoemission intensity from a localized open shell is caused by the orbital symmetry and polarization of the shell from which the electron is emitted. The angular distribution for non-resonant photoemission can thus be predicted from the dipole character of the process. The photoionization of a $d$-orbital yields a photoelectron with angular momentum as a superposition of $p$ and $f$ waves. The experimental value for the valence band is fairly close to the calculations of atomic Ni $3d$ orbitals by Yeh and Lindau [28] ($\beta_{PE}$ =1.145 at 800 eV).

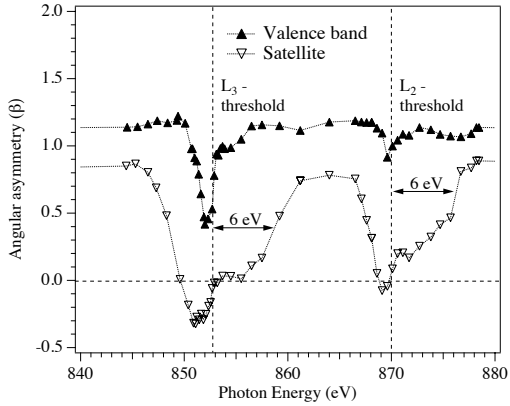

**Fig. 4:** Photon energy dependence of the asymmetry parameters for the valence band and the 6-eV satellite. In particular, for the satellite the negative β-parameters below each threshold are significant.

For the satellite, the angular intensity distribution is found to be lower than for the valence band with a $\beta_{PE}$-value of about 0.76 (dashed curve). It is generally known that electron-correlation effects play an important role for the 6-eV satellite. The photoemission spectrum of nickel metal has therefore often been discussed in terms of localized configurationally mixed states built from atomic configurations [29]. Thus, the difference in $\beta_{PE}$ for the valence band and the satellite probably reflects the difference in the symmetries of the excited states.

Figure 4 shows the variation of the $\beta_{PE}$ parameters as a function of photon energy for the valence-band and the satellite around the 2p thresholds. The thresholds corresponds to the XPS binding energies. The $\beta_{PE}$ values were obtained by using equation (1) under the assumption that the Auger signal is isotropic far away from the resonances, see above. Thus, in Fig. 4 the measured $\beta_{PE}$-values of the valence band and the 6-eV satellite in the fcc Ni crystal are displayed. The $\beta_{PE}$ parameters have been evaluated from the two extreme geometries. The angular distributions are found to experience strong oscillations through the resonances as the photon energy is increased. The variations are largest for the satellite while the valence band is somewhat less affected. From Fig. 4 it can be seen that the Fano-like resonance profiles are reflected also in $\beta_{PE}$. The angular asymmetry parameter $\beta_{PE}$ and the Fano asymmetry strength parameter, q, are thus alternative manifestations of the resonant behaviour. Starting with the valence band, it is clearly observed in Fig. 4 that the angular distribution is very different if the excitation energy is resonant or nonresonant. Below the $L_3$ resonance, the $\beta_{PE}$-value starts from about 1.15 and gradually decreases into a sharp minimum with a value of about 0.15. A similar but less pronounced behaviour is observed at the $L_2$ resonance. For the satellite, the effect is found to be much stronger with a decrease of the $\beta_{PE}$-parameter into a large asymmetric dip. In the nonresonant case, when the Auger matrix elements are not involved, the $\beta_{PE}$-value is about 0.76. Closely above the $L_3$ and $L_2$ thresholds, the angular distribution of the satellite is nearly isotropic as would be expected for a pure Auger decay. A negative $\beta_{PE}$-value of the satellite intensity is significantly found below both thresholds. In the transition region (0-6 eV) above each threshold, there is a mixture of the satellite photoemission and Auger matrix elements which is quite different from the behaviour of the angular distribution of the valence band. Thus, in the case of the satellite, the angular distribution strongly depends on if the Auger matrix elements are involved or not.

The interference effects are manifested in a pronounced decrease of the $\beta_{PE}$-values just below the thresholds, both for the valence band and the satellite. If we would assume that the spectra are incoherent superpositions of the photoemission and Auger signals, the observed β-values would be given as weighted mean values of the β-values for the individual processes;

$$\beta_{obs} = \frac{\beta_{PE} \times \sigma_{PE} + \beta_A \times \sigma_A}{\sigma_{PE} + \sigma_A} \qquad (2)$$

The photoemission intensity can be assumed to be constant over the resonance. Since $\beta_A = 0$, $\beta_{obs} \propto C/(\sigma_{PE}+\sigma_A)$, where C is a constant. If the process would be an incoherent superposition of the nonresonant photoemission signal with a $\beta_{PE}$-value of 1.15, with the isotropic Auger signal, the weighted mean average with the cross sections must always be positive. However, for the satellite intensity, a negative $\beta_P$-value is found below each resonance which can only be interpreted in terms of interference effects. It is attributed to the significant destructive interference in the 'photoemission geometry' at these photon energies [11]. The complex energy dependence of the $\beta_{PE}$-parameter clearly shows that in the case of Ni, there are not separated regimes with fixed predetermined values corresponding to a pure $cos^2\tau$ and a pure isotropic angular dependence, in the non-resonant and the resonant cases, respectively, as recently assumed in other publications [13]. Instead, the large difference in the angular distribution across the $L_{2,3}$ resonances and the large asymmetry of the dips clearly show that the giant resonant photoemission enhancement of the 6-eV satellite is due to a coherent process involving interference terms. The behavior of the asymmetry parameter above each resonance can be quantitatively described by assuming the incoherent superposition in Eq. 2. However, this data alone does not allow the possibility to exclude the presence of some coherent fraction.

# 4 Conclusions

The angular asymmetry and energy dependence of the resonant photoemission process in nickel metal at the 2p thresholds have been investigated. A true resonant photoemission phenomenon is understood as an interference between the direct photoemission channel followed by an Auger decay leading to the same final states when the photon energy is close to the absorption threshold. The massive intensity increase of the 6-eV correlation satellite at threshold is found to be at least two orders of magnitude. A smaller value of the $\beta_{PE}$-parameter describing the angular distribution is found for the satellite than for the valence band since the dynamical changes over the thresholds is much greater due to its different final state. The asymmetric decrease of the angular distribution with negative $\beta_{PE}$-values below each threshold provides unambiguous evidence that interference terms are involved in these energy regimes which is characteristic of a coherent resonant photoemission process.

# 5 Acknowledgments

This work was supported by the Swedish Natural Science Research Council (NFR), the Göran Gustavsson Foundation for Research in Natural Sciences and Medicine and the Swedish Institute (SI). ALS is supported by the U.S. Department of Energy, under contract No. DE-AC03-76SF00098.

# References

[1] C.-O. Almbladh and L. Hedin, *Handbook on Synchrotron Radiation*, Vol. 1, edited by E. E. Koch, (North-Holland, Amsterdam, (1983), and references therein.
[2] V. Schmidt, *Electron Spectrometry of Atoms using Synchrotron Radiation*, Cambridge University Press, 1997.
[3] N. Mårtensson, B. Johansson, Phys Rev. B **28**, 3733 (1983).

[4]     E. Antonides, E. C. Janse, and G. A. Sawatzky, Phys. Rev. B **15**, 1669 (1977); **15**, 4596 (1977).
[5]     G. van der Laan, M. Surman, M. A. Hoyland, C. F. J. Flipse, B. T. Thole, Y. Seino, H. Ogasawara and A. Kotani, Phys. Rev. B **46**, 9336 (1992).
[6]     O. Björneholm, J. N. Andersen, C. Wigren, A. Nilsson, R. Nyholm and N. Mårtensson, Phys. Rev. B **41**, 10408 (1990).
[7]     L. H. Tjeng *et al.*, Phys. Rev. Lett. **67**, 501 (1991).
[8]     C. Guillot *et al.*, Phys. Rev. Lett. **39**, 1632 (1977).
[9]     U. Fano, Phys. Rev. Lett **124**, 1866 (1961).
[10]    T. Kinoshita, T. Ikoma, A. Kakizaki, T. Ishii, J. Fujii, H. Fukutani, K. Shimada, A. Fujimori, T. Okane and S. Sato, Phys. Rev. B **47**, 6787 (1993).
[11]    M. Weinelt, A. Nilsson, M. Magnuson, T. Wiell, N. Wassdahl, O. Karis, A. Föhlisch, N. Mårtensson, J. Stöhr and M. Samant, Phys. Rev. Lett. **78**, 967 (1997).
[12]    M. F. López, A. Gutiérrez, C. Laubschat and G. Kaindl, J. Electr. Spectr. **71**, 73 (1995).
[13]    A. Gutiérrez and M. F. López, Phys. Rev. B **56**, 1111 (1997).
[14]    M. F. López, C. Laubschat, A. Gutiérrez, E. Weschke, A. Höhr, M. Domke and G. Kaindl, Surf. Sci. **307-309**, 907 (1994).
[15]    M. F. López, C. Laubschat, A. Gutiérrez, A. Höhr, M. Domke, G. Kaindl and M. Abbate, Z. Phys. B **94**, 1 (1994).
[16]    M. F. López, A. Höhr, C. Laubschat, M. Domke and G. Kaindl, Eorophys. Lett., **20** 357 (1992).
[17]    M. F. López, C. Laubschat and G. Kaindl, Europhys. Lett. **23**, 538 (1993).
[18]    L. H. Tjeng, Europhys. Lett. **23**, 535 (1993).
[19]    N. Mårtensson, P. Baltzer, P. A. Brühwiler, J.-O. Forsell, A. Nilsson, A. Stenborg, and B. Wannberg, J. Electr. Spectr. **70**, 117 (1994).
[20]    D. A. Shirley, Phys. Rev. B **5**, 4709 (1972).
[21]    *Practical Surface Analysis by Auger and X-ray Photoelectron Spectroscopy*, edited by D. Briggs and M. P. Seah, John Wiley & Sons Ltd 1987
[22]    We note that based on the detailed analysis here, in Ref.[11] (Fig. 1) the relative normalization between the two orientations is different for some of the photon energies. However, this does not affect the quantitative conclusions.
[23]    W. Wurth, G. Rocker, P. Feulner, R. Scheuerer, L. Zhu and D. Menzel, Phys. Rev. B **47**, 6697 (1993).
[24]    J. Fuggle, P. A. Bennett, F. U. Hillebrecht, A. Lenselink and G. A. Sawatzky, Phys. Rev. Lett. **49**, 1787 (1982).
[25]    A. Sommerfeld and G. Schur, Ann. Phys. (Leipzig) **4**, 410 (1930).
[26]    J. Berkowitz, *Photoabsorption, Photoionization and Photoelectron Spectroscopy*, Academic Press 1979.
[27]    J. Cooper and R. N. Zare, *Lect. Theor. Phys.* **11C**, 323 (1969).
[28]    J. J. Yeh and I. Lindau, Atomic data and Nuclear Tables **32**, 2 (1985).
[29]    T. Jo and G. A. Sawatzky, Phys. Rev. B **43**, 8771 (1991).
[30]    B. Kämmerling, A. Hausmann, A. Länger and V. Schmidt, J. Phys. B **25**, 4773 (1992).